\documentclass[%
 reprint,
 amsmath,amssymb,
 aps,
]{revtex4-2}

\usepackage{hyperref}
\usepackage{graphicx}
\usepackage{dcolumn}
\usepackage{bm}
\usepackage[usenames, dvipsnames]{color}
\usepackage{color}
\usepackage{amsmath}
\usepackage{comment}

\begin{document}

\author{Roi Holtzman}
 \email{roi.holtzman@weizmann.ac.il}
\author{Oren Raz}%
 \email{oren.raz@weizmann.ac.il}
\affiliation{Department of Physics of Complex Systems, Weizmann Institute of Science, Rehovot, 76100, Israel}
\author{Christopher Jarzynski}%
 \email{cjarzyns@umd.edu}
 \affiliation{Institute for Physical Science and Technology, University of Maryland, College Park, Maryland 20742, USA; \\
   Department of Chemistry and Biochemistry, University of Maryland, College Park, Maryland 20742, USA \\
 and Department of Physics, University of Maryland, College Park, Maryland 20742, USA}

\date{\today}

\title{Shortcuts to adiabaticity across a separatrix}

\begin{abstract}

Shortcuts to adiabaticity are strategies for conserving adiabatic invariants under non-adiabatic (i.e.\ fast-driving) conditions.
Here, we show how to extend classical, Hamiltonian shortcuts to adiabaticity to allow the crossing of a phase-space separatrix -- a situation in which a corresponding adiabatic protocol does not exist.
Specifically, we show how to construct a time-dependent Hamiltonian that evolves one energy shell to another energy shell across a separatrix. 
Leveraging this method, we design an erasure procedure whose energy cost and fidelity do not depend on the protocol's duration.

\end{abstract}

\maketitle

At the first Solvay conference in 1911, Lorentz posed the question, ``How does a simple pendulum behave when the length of the suspending thread is gradually shortened?''~\cite{Littlewood_1963_LorentzPendulumProblem}.
The following day, Einstein responded that the pendulum's energy and frequency both change with time, but their ratio remains constant, provided the thread's length changes sufficiently slowly (and the amplitude of oscillations is small).
This example illustrates the notion of an {\it adiabatic invariant} --- a quantity that remains constant when the Hamiltonian is varied infinitely slowly.
For classical systems in one degree of freedom, which we consider in this paper, the {\it action}, $I = \oint p \, dq$, is an adiabatic invariant \cite{Lifshitz_1976_Mechanics}.
Note that the notion of adiabaticity discussed here differs from that used in thermodynamics where an adiabatic process is one that involves no heat exchange.

The adiabatic invariant is not an exact constant of motion, but its deviation from the initial value is small when the Hamiltonian changes on a timescale, $\tau$, that is much longer than one orbital period, $T_{\rm per}$, of the system's unperturbed motion.
Performing such slow protocols is often impractical, and a line of work that is now termed \emph{shortcuts to adiabaticity} (STA) \cite{Muga_2019_ShortcutsAdiabaticityConcepts} was developed for achieving the outcomes of adiabatic transformations without the restriction of slow manipulations of the Hamiltonian.
STA allow one to perform protocols rapidly, while guaranteeing that at the end of the protocol, the value of the adiabatic invariant has not changed.

The terms {\it counterdiabatic} (CD) and {\it fast-forward} (FF) refer to two prominent approaches for constructing STA.
Both involve the addition of an auxiliary Hamiltonian term, designed to conserve the value of the adiabatic invariant.
Under CD driving, the adiabatic invariant remains constant throughout the protocol, but the auxiliary term is a function of both position and velocity, which could be difficult to implement in the laboratory.
In the FF approach, the auxiliary term is a time-dependent potential-energy function, and the adiabatic invariant varies with time, but returns to its initial value at the end of the protocol \cite{Subasi_2017_FastForwardClassical, Jarzynski_2017_ShortcutsAdiabaticityUsing,Guery-Odelin2023}.

Here, we seek to extend classical STA to situations where the notion of adiabaticity breaks down.
Consider the task of evolving Lorentz's pendulum from a librating (``back-and-forth'') to a rotating (``round-and-round'') trajectory, while guaranteeing that the final value of the adiabatic invariant, $I$, is equal to its initial value. At the transition between libration and rotation the system crosses a separatrix, and at this energy its orbital period diverges.
As a result, the adiabaticity criterion ($\tau\gg T_{\rm per}$) cannot be satisfied, and the adiabatic invariance of the classical action, $I$, breaks down~\cite{Tennyson_1986_BreakdownOfAdiabaticInvariant}.
In this paper, we show how to design an STA protocol that successfully guides a trajectory from one side of the separatrix to the other while maintaining the value of the adiabatic invariant.

Crossing a separatrix demonstrates a useful application of STA ideas, where the standard approach of performing a protocol slowly is doomed to fail.
Additional motivation for crossing a separatrix arises naturally in the context of information erasure.
In recent years, much attention has been focused on approaching the Landauer bound \cite{Landauer_1961_IrreversibilityHeatGeneration} and minimizing the dissipated heat associated with erasing a bit of information \cite{Bechhoefer_2020_OptimalFinitetimeBit, DeWeese_2014_OptimalFinitetimeErasure, Crutchfield_2022_ShortcutsThermodynamicComputing,holtzman2021hamiltonian}.
A typical system that realizes a logical bit is a particle in a double-well potential, and erasing the bit implies mapping an initial distribution that is located in both wells into a final distribution that is located in a single well.
Performing such a protocol in the underdamped limit was recently analyzed theoretically and performed experimentally in \cite{Bellon_2021_InformationThermodynamicsFast, Bellon_2022_DynamicsInformationErasure, Bellon_2023_AdiabaticComputingOptimal}.
The main limitation of such erasure protocols is that typically, the faster the erasure is performed, the higher the energy needed for the operation.
We show that by crossing the separatrix while keeping the adiabatic invariant unchanged, it is possible to implement an erasure protocol such that neither its energy cost nor its fidelity is affected by the rate of the protocol.

In what follows, our general discussion is illustrated by a specific example: a double-well potential. 
An additional example -- the pendulum -- is included in Appendix \ref{sec:pendulum-appendix}.
Both examples yield analytic expressions for the CD and FF auxiliary terms.
On the separatrix, these terms diverge at the energy shell's fixed points.
However, a proper choice of the protocol $\lambda(t)$ allows one to regularize these divergences, as we demonstrate using numerical simulations.  
In our approach, we constrain this protocol to start and end at given values, $\lambda_0 \equiv \lambda(0)$ and $\lambda_\tau \equiv \lambda(\tau)$, while allowing ourselves freedom in the functional form $\lambda(t)$ interpolating from $\lambda_0$ to $\lambda_\tau$.
A proper choice of this interpolation is crucial for conserving the adiabatic invariant across the separatrix.

Consider a particle in a one-dimensional double-well potential described by the Hamiltonian
\begin{equation}
\label{eq:double-well}
H(q, p, \lambda) = \frac{p^2}{2} + \frac{q^4}{4} - \frac{\lambda^2 q^2}{2}
\equiv \frac{p^2}{2} + U(q, \lambda) \, ,
\end{equation}
where \(q, p\) are the coordinate and momentum, respectively, and \(\lambda\) is a control parameter that determines the depth of the wells and the position of the minima.
For simplicity, the particle's mass is set to unity.

\begin{figure*}[htbp]
  \centering
  \includegraphics[width=0.99\textwidth]{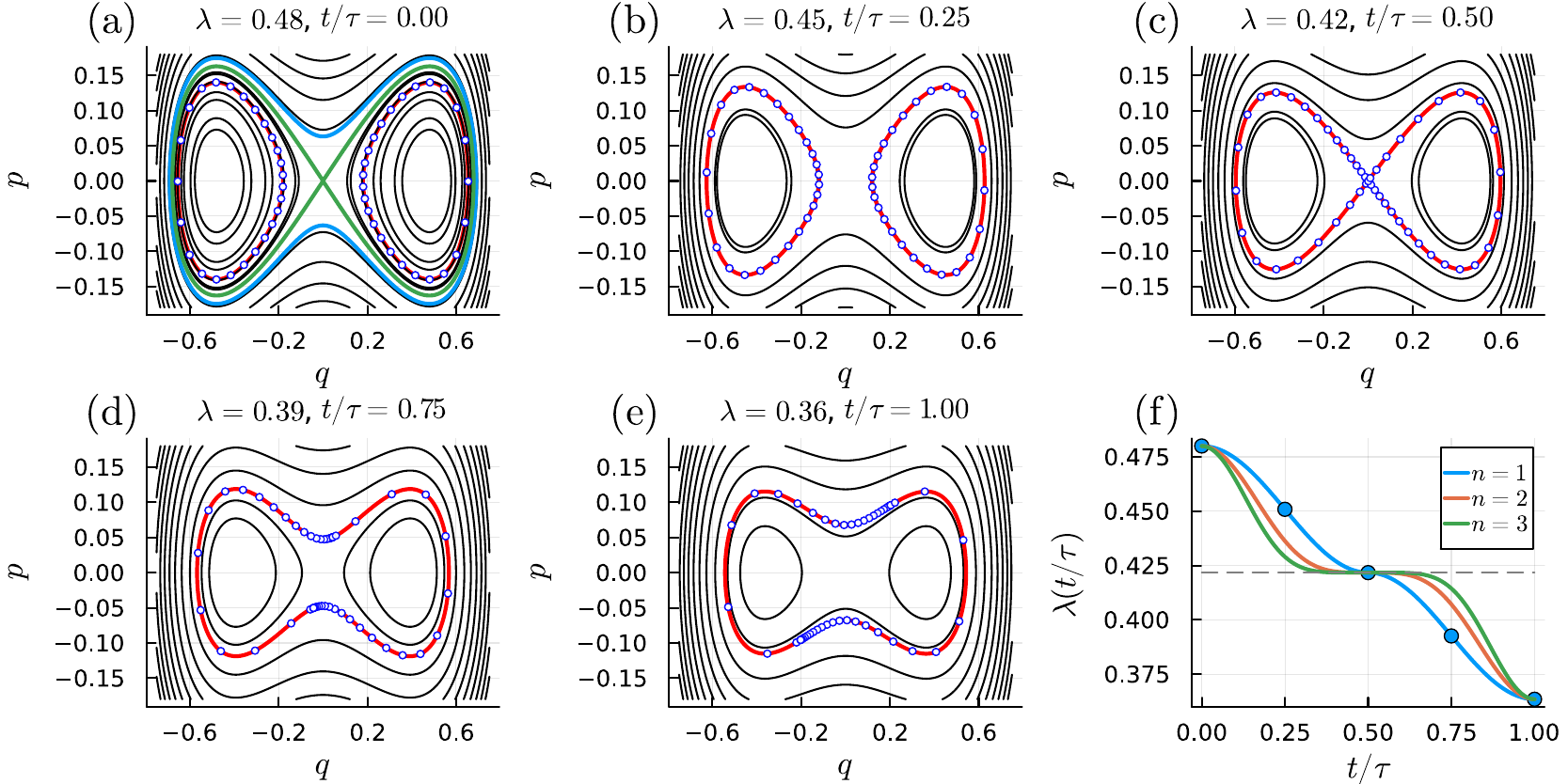}
  \caption{(a-e) The evolution of an energy shell crossing a separatrix with CD driving using the protocol $\lambda(t)$ in panel (f). The five points in panel (f) mark the times of the snapshots shown in (a-e). We sample 100 initial conditions (blue circles) microcanonically on a particular energy shell $E_0$ at $\lambda_0$. From each initial condition a trajectory evolves under the Hamiltonian $H_{\rm tot}(t) = H(\lambda(t)) + H_{CD}(t)$ with $\lambda(t)$ given by Eq.~\eqref{eq:lambda-t} for $n=1$ (panel (f)). We see that every trajectory remains on the instantaneous energy shell throughout the process, crucially also crossing the separatrix. In panel (a), we highlight energy shells above, on, and below the separatrix in blue, green, and red, respectively. Panel (f) shows the protocol $\lambda(t)$ for $n=1,2,3$. Note that the approach of $\dot{\lambda}(t) \to 0$ as $t \to t_s = \tau/2$ is stronger as $n$ increases.} 
  \label{fig:shell-evolution-CD}
\end{figure*}

For a given value of $\lambda$, the {\it energy shell} with energy $E$ is the set of all points $(q,p)$ such that $H(q,p,\lambda)=E$. For any $\lambda \neq 0$, our double-well system has a special energy shell, the {\it separatrix},  of energy $E_s = 0$.
The separatrix divides phase space into two regions characterized by different energy-shell topologies:
every energy shell with $E \in [- \lambda^{4} / 4, 0)$ is composed of two ergodic components (two closed loops in phase space), each residing in one well, whereas every energy shell with $E \in \left( 0, \infty \right)$ forms a simple closed loop that encompasses both wells. The separatrix $E_s=0$ forms a figure-eight curve that separates these topologically distinct families of energy shells, as highlighted by the 
red, green, and blue lines in Fig.~\ref{fig:shell-evolution-CD}(a).

Consider initial values $\lambda_0 \neq 0$ and $E_0 < 0$. Our goal is to vary $\lambda$ with time to a given final value $\lambda_{\tau}$, such that all initial conditions on the energy shell $E_0$ are mapped --- under Hamiltonian time-evolution from $t=0$ to $t=\tau$ --- to the same final energy shell $E_{\tau} >0$.
By construction, every such trajectory crosses the separatrix \footnote{The choice of going from below the separatrix to above the separatrix is taken only for the sake of concreteness.
The opposite transition is performed in the same manner.}.  The value of $E_\tau$ is set by Liouville's theorem: if such a protocol exists, the phase-space volumes enclosed by the two energy shells must be identical.
Thus we have
\begin{equation}
    \Omega(E_0,\lambda_0) = \Omega(E_\tau,\lambda_\tau) \, ,
\end{equation}
where $\Omega(E, \lambda) = \oint_{H(q, p, \lambda) = E} p dq$ is the volume enclosed by the energy shell $E$ of $H(q,p,\lambda)$.
For $\Omega_0\equiv\Omega(E_0,\lambda_0)$, we define the {\it adiabatic energy} $E_{\Omega_0}(\lambda)$ by the condition
\begin{equation}
\label{eq:adiabaticEnergy}
    \Omega( E_{\Omega_0}(\lambda), \lambda ) = \Omega_0 \, .
\end{equation}
We want to construct a protocol under which the time-dependent energy of every trajectory evolving from the initial energy shell $E_0$, is the instantaneous adiabatic energy, $E_{\Omega_0}(\lambda(t))$.

Due to the above-mentioned breakdown of the adiabatic invariant when the system crosses a separatrix, 
our goal cannot be accomplished simply by varying $\lambda(t)$ slowly.
Intuitively, under adiabatic driving, all trajectories on a given instantaneous energy shell evolve for many periods of oscillation around that shell before $\lambda$ changes significantly.
This observation underlies the idea of adiabatic averaging: the effect of the time-dependence of $\lambda$ is approximately the same for all trajectories on a given energy shell, and is determined by a microcanonical average computed over that shell~\cite{Lifshitz_1976_Mechanics}.
As the parameter is varied slowly from $\lambda$ to a slightly different value $\lambda+\Delta\lambda$, all trajectories on the energy shell $E_{\Omega_0}(\lambda)$ evolve to the energy shell $E_{\Omega_0}(\lambda+\Delta\lambda)$, due to adiabatic averaging.
At the separatrix, however, the energy shell has a fixed point, and the time required for a trajectory to explore the entire energy shell diverges.
As a result, adiabatic averaging breaks down and the intuitive argument for the adiabatic theorem fails.

In the CD approach, one adds an auxiliary Hamiltonian term (Eq.~\ref{eq:H_CD} below) whose effect is to evolve any point on the energy shell $E_{\Omega_0}(\lambda)$ to the energy shell $E_{\Omega_0}(\lambda+\Delta\lambda)$ when the parameter is varied from $\lambda$ to $\lambda+\Delta\lambda$, regardless of the driving rate $\dot\lambda$~\cite{Subasi_2017_FastForwardClassical, Jarzynski_2017_ShortcutsAdiabaticityUsing}.
As this approach does not rely on adiabatic averaging, one might suppose that CD driving can be used to achieve our goal of keeping the adiabatic invariant fixed while driving the system across a separatrix.
However, the CD approach generically fails at this task.
In what follows, we discuss why it fails, before showing that it can be ``fixed'' by a proper choice of the protocol $\lambda(t)$ \footnote{
Details on the FF driving are included in the appendix \ref{sec:app-dw-acc-field}.}.

The auxiliary, or counterdiabatic, term that one adds to the original Hamiltonian $H(\lambda(t))$ is
\begin{equation}
\label{eq:H_CD}
    H_{CD}(t) = p \tilde{v}(q, t) = p v(q, \lambda(t)) \dot\lambda(t),
\end{equation} 
where $v(q, \lambda)$ is a ``velocity field''\footnote{The reason that $v(q, \lambda)$ is named a ``velocity field'' is due to the manner in which this function is constructed in Ref. \cite{Subasi_2017_FastForwardClassical}.
There, an energy shell is partitioned to equal area segments using a set of points $\left\{ q_i \right\}_{i=1}^N$.
Changing the value of $\lambda$ alters the shell, and so the set of points $\left\{ q_i \right\}$ are moved to maintain their partition of equal areas.
This motion of points is what defines the function $v(q, \lambda)$. See Appendix \ref{sec:app-v-def}.
Similarly, a corresponding acceleration field $a(q, \lambda)$, which is given by appropriate derivatives of $v(q, \lambda)$, can be defined.} given by \cite{Subasi_2017_FastForwardClassical, Jarzynski_2017_ShortcutsAdiabaticityUsing}
\begin{align}
\label{eq:v-def}
v(q, \lambda) &= - \frac{1}{\overline{p}} \frac{\partial }{\partial \lambda}  \int_{q_{l}(\lambda)}^q \overline{p}(q', \lambda) dq'\\
\label{eq:pbar-def}
  \overline{p}(q, \lambda) &= \sqrt{2 \left( E_{\Omega_0}(\lambda) - U(q, \lambda) \right)}.
\end{align}
The functions $v(q,\lambda)$ and $\overline{p}(q,\lambda)$ depend on the chosen initial values $\lambda_0$ and $E_0$ via the adiabatic energy $E_{\Omega_0}(\lambda)$, which is set by Eq.~\eqref{eq:adiabaticEnergy}.
In Eq.~\eqref{eq:v-def}, $q_{l}(\lambda)$ is the left turning point of the energy shell $E_{\Omega_0}(\lambda)$.
If there is no turning point (as in the librational motion of the pendulum) $q_{l}(\lambda)$ can be assigned arbitrarily (see details in appendix \ref{sec:app-v-def}).

Equation \eqref{eq:v-def} defines $v(q, \lambda)$ only at values of $q$ satisfying $U(q,\lambda)<E_{\Omega_0}(\lambda)$, i.e.\ ``inside'' the energy shell $E_{\Omega_0}(\lambda)$.
We extend this definition to include turning points $q_{\rm tp}$ at the edge of the energy shell, where $U(q,\lambda)=E_{\Omega_0}(\lambda)$, by setting (see appendix \ref{sec:app-v-def})
\begin{equation}
\label{eq:v-def-turning-point-limit}
v_{q_{\rm tp}}(\lambda) = \lim_{q \to q_{\rm tp}} v(q, \lambda),
\end{equation}
where $q$ approaches $q_{\rm tp}$ from within the energy shell.
This limit coincides with the velocity with which the turning points move as the control parameter $\lambda$ is changed: $v_{q_{\rm tp}} = d q_{\rm tp}/d \lambda$ (see appendix \ref{sec:app-v-def}).
By construction~\cite{Subasi_2017_FastForwardClassical}, trajectories with initial energy $E_0$ remain on the energy shell $E_{\Omega_0}(\lambda(t))$ at all times $t$, therefore there is no need to define the velocity field for values of $q$ that are outside the shell ($U(q,\lambda)>E_{\Omega_0}(\lambda)$).

While $v(q,\lambda)$ is finite at the turning points, it generically diverges at the fixed point of a separatrix.
To see this, note that as $\lambda$ is varied, a fixed point appears exactly when two ergodic components touch each other, forming a figure-eight curve as two turning points coalesce. 
The implicit equation defining a turning point,
\begin{equation}
    U\left( q_{\rm tp}(\lambda) , \lambda \right) - E_{\Omega_0} \left(\lambda \right) = 0,
\end{equation}
yields the turning-point velocity
\begin{equation}
\label{eq:v-turning-point}
    v_{q_{\rm tp}} = \frac{d q_{\rm tp}}{d \lambda} = \frac{\partial_{\lambda} \left( E_{\Omega_0}(\lambda) - U(q_{\rm tp}, \lambda) \right)}{\partial_q U(q_{\rm tp}, \lambda)}.
\end{equation}
Since $-\partial_q U(q_{\rm tp}, \lambda)$ is the force the potential exerts on the particle at the turning point, the denominator in Eq.~\eqref{eq:v-turning-point} is nonzero at an ordinary turning point, but vanishes at the fixed point of a separatrix, where two turning points meet at the point where the potential has a maximum.
As there is no generic reason for the numerator to vanish at the fixed point, $v_{q_{\rm tp}}$ generally diverges there.
This conclusion is supported by explicit calculations of $v(q, \lambda)$ for the double-well (see appendix \ref{sec:app-v-at-turning-points}) and the pendulum (see appendix \ref{sec:pendulum-appendix}).

\begin{figure}[htbp]
  \centering
  \includegraphics[width=0.99\columnwidth]{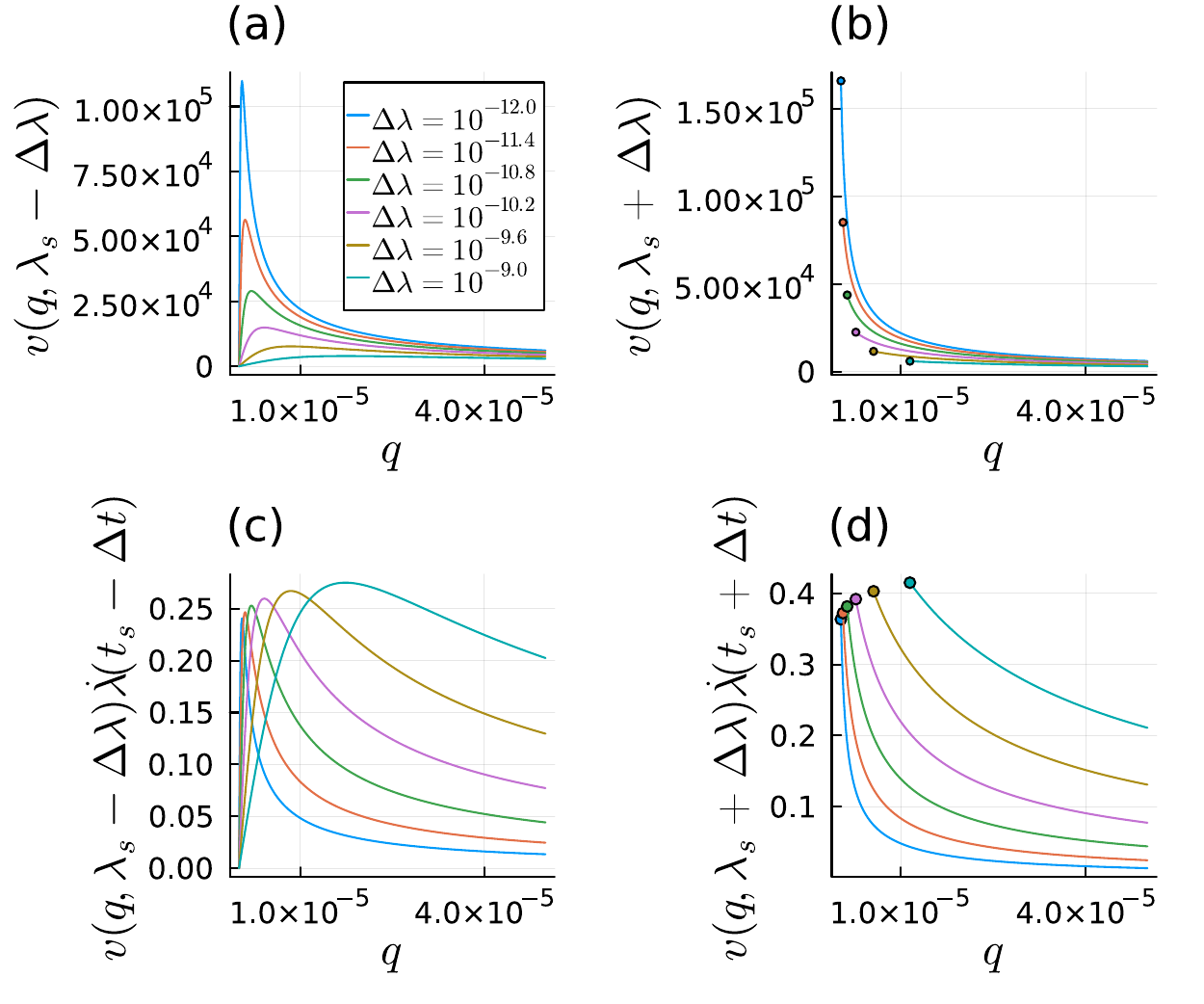}
  \caption{The velocity field $v(q, \lambda)$ (top panel) and the product $\tilde{v}(q,t)=v(q, \lambda(t)) \dot{\lambda}(t)$ (bottom panel) are plotted at values $\lambda = \lambda_s \pm \Delta \lambda$.
  The left and right panels correspond to approaching the separatrix from above and below, respectively. 
  The legend in panel (a) corresponds to all panels.
  In the bottom panels (c,d), the protocol $\lambda(t)$ of Eq.~\eqref{eq:lambda-t}, with $n=1$, is used.
  The circles in (b,d) are the inner turning points $q_{\rm tp}^*$, marking the edge of the energy shell where $v(q, \lambda)$ is defined (see text).
  Note the great difference in magnitudes between the vertical scales in the top panel and those in the bottom panel.
  These plots demonstrate both that $v(q, \lambda)$ diverges upon approaching the separatrix, and that this divergence can be restrained in $\tilde{v}(q, t)$ by taking the rate of the approach of $\lambda(t \to t_s)$ to $\lambda_s$ to be small enough.   }
  \label{fig:v_of_q}
\end{figure}

The top panel of Fig.~\ref{fig:v_of_q} illustrates the divergence in the velocity field $v(q,\lambda)$ that occurs at the fixed point of the separatrix, for the double-well potential of Eq.~\eqref{eq:double-well}.
For a given choice of $\lambda_0$ and $E_0<0$, let $\lambda_s$ denote the value of $\lambda$ at which the separatrix is crossed, defined by the condition $E_{\Omega_0}(\lambda_s)=0$.
The separatrix has a fixed point at $q=0$.
For $\lambda>\lambda_s$, the adiabatic energy shell $E_{\Omega_0}(\lambda)<0$ has two lobes, placed symmetrically around the fixed point.
The left turning point of the right lobe, and its counterpart on the left lobe, are located at $\pm q_{\rm tp}^*(\lambda)$, and they converge to the fixed point as $\lambda \rightarrow \lambda_s$.
For six small, positive values $\Delta\lambda$, Fig.~\ref{fig:v_of_q}(b) plots $v(q,\lambda_s+\Delta\lambda)$ as a function of $q$, over a small region to the right of the fixed point.
We see that as $\Delta\lambda$ decreases from $10^{-9}$ to $10^{-12}$, $v(q_{\rm tp}^*,\lambda)$ increases rapidly. 
For $\lambda<\lambda_s$, the adiabatic energy shell $E_{\Omega_0}(\lambda)>0$ forms a single, closed loop, which is symmetric around $q=0$.
Figure~\ref{fig:v_of_q}(a) plots $v(q,\lambda_s-\Delta\lambda)$, for the same values of $\Delta\lambda$ and over the same range of $q$ as in Fig.~\ref{fig:v_of_q}(b).
While $v(q,\lambda)$ vanishes identically at $q=0$, for small $\Delta\lambda$ a peak forms in the velocity field.
The location of this peak approaches the fixed point, and its magnitude increases rapidly, as $\Delta\lambda$ decreases.
The growth of $v(q_{\rm tp}^*,\lambda)$ in Fig.~\ref{fig:v_of_q}(b), and of the peak size in Fig.~\ref{fig:v_of_q}(a), reflect the divergence in $v(q,\lambda_s)$ at the fixed point $q=0$.

The divergence of $v(q,\lambda)$ at the fixed point of a separatrix may cause the CD Hamiltonian $H_{CD} = p v(q, \lambda) \dot{\lambda}$ (see Eq.~\eqref{eq:H_CD}) to become ill-defined at the separatrix. However, the form of $H_{CD}$ suggests a solution: we can suppress the effect of a divergent $v(q,\lambda)$ by choosing $\lambda(t)$ such that the product $v(q, \lambda)\dot{\lambda}$ remains finite.
This requires a protocol $\lambda(t)$ that has an inflection point exactly as the system crosses the separatrix.
We choose a protocol, from $t=0$ to $\tau$, of the form
\begin{equation}
\label{eq:lambda-t}
\lambda(t) = \begin{cases}
                   \lambda_s +  \left( \lambda_0 - \lambda_s \right) \cos^{2n} \left( \pi t / \tau \right) , & t \le \tau / 2 \\
                    \lambda_s +  \left( \lambda_{\tau} - \lambda_s \right)  \cos^{2n} \left( \pi t / \tau \right) , & t > \tau / 2,
                 \end{cases}
\end{equation}
where $n>0$ is an integer.
Under this protocol, the separatrix is crossed at $t_{s} = \tau / 2$, and $\dot{\lambda}(t_s)=0$.
As $t \to t_s$, $\dot\lambda(t)$ scales as $-\vert t-t_s\vert^{2n-1}$, as illustrated in Fig.~\ref{fig:shell-evolution-CD}(f).
By choosing $n$ to be sufficiently large, we may be able to regulate the divergence of $v(q, \lambda \to \lambda_s)$.
For convenience we henceforth set $\lambda_{\tau} = 2\lambda_s - \lambda_0$, so that the protocol is symmetric around the separatrix: $\lambda_s = (\lambda_0+\lambda_\tau)/2$.

The bottom panel of Fig.~\ref{fig:v_of_q} illustrates how the divergence in $v(q,\lambda)$ is suppressed in the product $\tilde{v}(q,t)=v(q, \lambda(t)) \dot{\lambda}(t)$, for the protocol $\lambda(t)$ (Eq.~\eqref{eq:lambda-t}) with $n=1$.
Figs.~\ref{fig:v_of_q}(d) and (c) plot $\tilde v(q,t)$ at times $t=t_s-\Delta t$ and $t=t_s+\Delta t$, respectively, where $\lambda(t_s\mp\Delta t) = \lambda_s\pm\Delta\lambda$, for the same values of $\Delta\lambda$ as in the top panel.
We observe no divergent behavior either in the value of $\tilde v(q_{tp}^*,t)$ in Fig.~\ref{fig:v_of_q}(d), or in the size of the peak in $\tilde v(q,t)$ in Fig.~\ref{fig:v_of_q}(c), as $\lambda$ approaches $\lambda_s$.

To demonstrate the applicability of our protocol, we evolve Hamilton's equations of motion for $H_{\rm tot}(t) = H + H_{CD}$ given by Eqs.~\eqref{eq:double-well} \eqref{eq:H_CD}, with the protocol $\lambda(t)$ for $n=1$ in Eq.~\eqref{eq:lambda-t} (see Fig.~\ref{fig:shell-evolution-CD}(f)).
We chose values $\lambda_0=0.48$ and $\Omega_0 = 0.2$ corresponding to an energy shell below the separatrix.
100 initial conditions were sampled microcanonically on the shell and were evolved under $H_{\rm tot}(t)$.
Snapshots of the evolution are shown in Fig.~\ref{fig:shell-evolution-CD}. We observe that all points follow the instantaneous energy shell, even across the separatrix.

The idea of crossing a separatrix can be used to erase a bit.
Recently, the realization of a bit and its erasure using an underdamped particle in a double-well was implemented experimentally~\cite{Bellon_2021_InformationThermodynamicsFast, Bellon_2022_DynamicsInformationErasure, Bellon_2023_AdiabaticComputingOptimal}.
The erasure procedure has three steps: (a) Merge the two wells, (b) Translate the single well to, say, the right, and (c) Recreate another empty well on the left (see Fig. (1) in Ref.~\cite{Bellon_2023_AdiabaticComputingOptimal}).
The Merge step is essentially analogous to crossing a separatrix as it joins energy shells that are composed of two disjoint loops into a single loop.
The limit of zero damping in such underdamped systems corresponds to an isolated Hamiltonian system, which is the context of the present paper.

\begin{figure}[htbp]
  \centering
  \includegraphics[width=1\columnwidth]{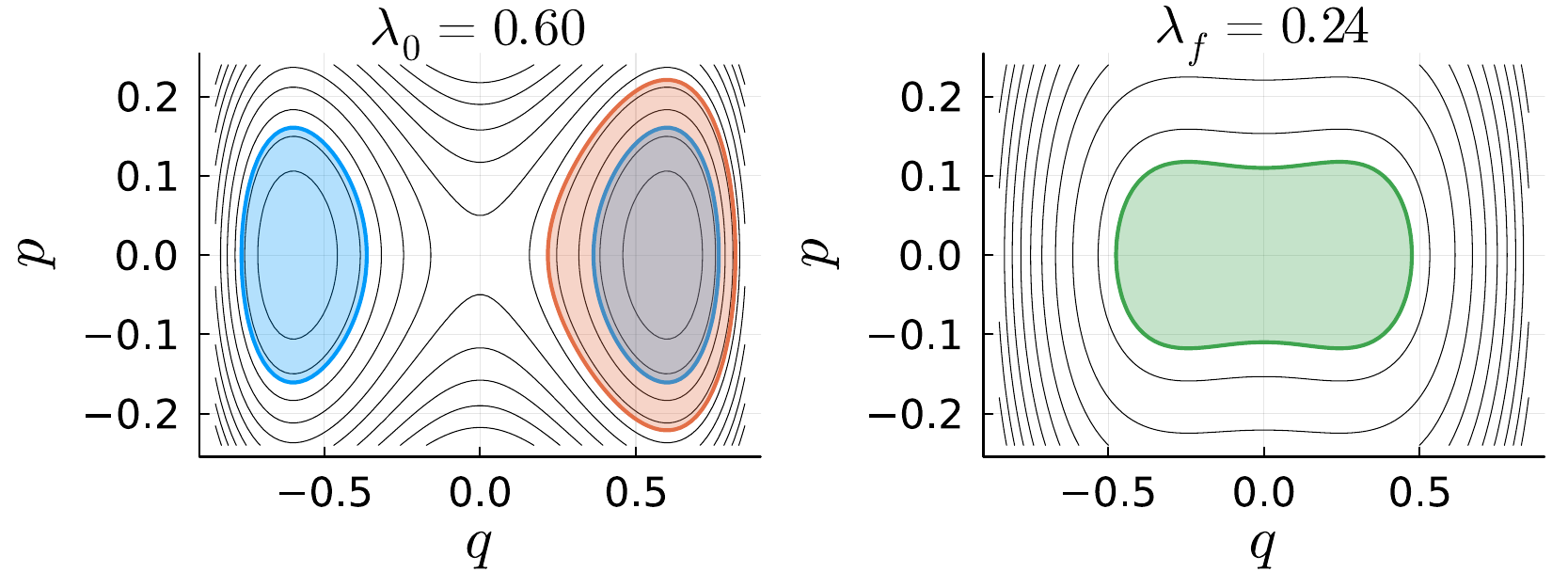}
  \caption{Erasure protocol for the double-well in phase space. The system begins in equilibrium, at a sufficiently low temperature that the distribution is effectively confined inside an energy shell $E_0$, depicted by the pair of blue loops in the left panel. Then, a CD protocol taking $\lambda_0 \mapsto \lambda_f$ maps the blue energy shell (below the separatrix) to the green energy shell (above the separatrix). Next, another CD protocol taking $(\lambda_f, \alpha=0) \mapsto (\lambda_0, \alpha=0)$, see Eq.~\eqref{eq:double-well-alpha}, maps the green energy shell to the orange contour on the left panel corresponding to the right lobe of an energy shell $E_r$ at $\lambda_0$. Since the entire evolution depicted here is Hamiltonian, the blue, green and orange contours enclose equal volumes, by Liouville's theorem.}
  \label{fig:erasure}
\end{figure}

We now propose a specific erasure protocol for our double-well system (Eq.~\eqref{eq:double-well}), inspired by Refs.~\cite{Bellon_2021_InformationThermodynamicsFast, Bellon_2022_DynamicsInformationErasure, Bellon_2023_AdiabaticComputingOptimal}, that leverages the method we have developed for crossing a separatrix.
The proposed protocol can be implemented rapidly, and neither its energetic cost nor fidelity scale with the protocol duration.
Initially, the system is connected to a heat bath at temperature $T$, which is sufficiently low that the equilibrium distribution is concentrated within the wells, with $\lambda$ set at a value $\lambda_0$.
We can then specify an energy shell $E_0<0$, determined by $T$, that effectively bounds the distribution: the probability that the system's energy exceeds $E_0$ is vanishingly small.
The two blue loops in Fig.~\ref{fig:erasure} indicate this energy shell.
The system is now disconnected from the bath, hence its evolution in the next stages is governed by Hamilton's equations.

Next, we implement the CD driving for separatrix crossing developed above.
The control parameter is varied from $\lambda_0$ to $\lambda_\tau$, causing points on the energy shell $E_0<0$ (blue loops in Fig.~\ref{fig:erasure}) to evolve to an energy shell $E_f>0$ above the separatrix (green loop).
Since Hamiltonian trajectories cannot cross in phase space, the region of phase space enclosed by the energy shell $E_0$ in the left panel in Fig.~\ref{fig:erasure} is mapped to the region enclosed by the energy shell $E_f$ in the right panel.

The next step is to evolve the energy shell $E_f$ at $\lambda_f$ (green loop) to the right lobe of the energy shell $E_r$ at $\lambda_0$ (orange loop).
By Liouville's theorem, the volume enclosed by this lobe is equal to the total volume enclosed by the energy shell $E_0$.
That is, $\Omega(E_{r}, \lambda_0)/2 = \Omega(E_0, \lambda_0)$.
This last transformation can be achieved by including an additional linear term in the potential,
\begin{equation}
\label{eq:double-well-alpha}
    U(q, \lambda, \alpha) = \frac{q^4}{4} - \frac{\lambda^2 q^2}{2} - \alpha q,
\end{equation}
and choosing a protocol $(\lambda(t),\alpha(t))$ from $(\lambda_f,0)$ to $(\lambda_0,0)$ that smoothly deforms the green loop to the orange loop.
Since this transformation involves no separatrix crossings, it can be implemented using previously developed STA methods~\cite{Subasi_2017_FastForwardClassical, Jarzynski_2017_ShortcutsAdiabaticityUsing}.

The last step is to return the system to its initial distribution, effectively compressing the orange phase space volume to the right lobe of the blue volume in Fig.~\ref{fig:erasure}. 
This is accomplished by reconnecting the system to the bath and letting it equilibrate. 
The energy that was injected into the system during the protocol now dissipates to the bath as heat, as the system relaxes to its original distribution.

The protocol taking the initial distribution that resides in both wells at $\lambda_0$ (blue area in Fig.~\ref{fig:erasure}) to the right well at the same $\lambda_0$ (orange area) can be done rapidly (since it leverages the methods of STA), and the energy cost is bounded by the constraint of Liouville's theorem. 
The last step, in which the system is reconnected to the bath at temperature $T$ and relaxes back to equilibrium, is governed by the coupling strength between the bath and the system.
Controlling the rate of this step is out of the scope of this letter.

In conclusion, in this letter, we have shown that classical shortcuts to adiabaticity can be modified to apply even when the notion of adiabaticity breaks down due to the existence of a separatrix.
Direct implementation of CD driving fails at the separatrix as $H_{CD}$ diverges at the fixed point. 
This divergence can be regularized by choosing a protocol $\lambda(t)$ that has an inflection point when crossing the separatrix. 
The same conclusions apply to FF driving (see Appendix \ref{sec:app-dw-acc-field}).
The method of crossing a separatrix can be used in practical applications such as the fundamental task of information erasure.

\paragraph*{Acknowledgements:} OR acknowledges financial support from the ISF, grant no. 232/23, and of the Minerva Stiftung. 
CJ acknowledges support by the U.S. National Science Foundation under Grant No. 2127900. RH acknowledges support from the U.S.-Israel Binational Science Foundation under the Prof. Rahamimoff Travel Grants for Young Scientists.

\bibliographystyle{unsrt}
\bibliography{STA}

\newpage

\appendix

\section{Definitions of the velocity field}
\label{sec:app-v-def}

Here we define the velocity field $v(q,\lambda)$, Eq.~\eqref{eq:v-def}, following Refs. \cite{Subasi_2017_FastForwardClassical, Jarzynski_2017_ShortcutsAdiabaticityUsing}.
For a Hamiltonian $H(q,p,\lambda_0)$, consider an energy shell $E_0$ that forms a simple closed loop in phase space, enclosing a volume $\Omega_0 = \Omega(E_0, \lambda_0)$.
Take a set of $N$ points $\{q_i\}_{i=1}^{N}$ such that vertical lines passing through them (i.e., perpendicular to the $q$-axis) partition the energy shell into $N+1$ strips of equal phase-space volume.
The volume enclosed by the energy shell, up to the point $q_i$, is given by
\begin{equation}
\label{eq:S-qi}
    S(q_i, \lambda_0) = 2 \int_{q_l(\lambda_0)}^{q_i} \overline{p} \, dq,
\end{equation}
where $\overline{p}(q, \lambda) = \sqrt{2 \left( E_{\Omega_0}(\lambda) - U(q, \lambda) \right)}$ and $q_l(\lambda_0)$ is the left turning point of the energy shell.
Upon changing the value of the control parameter, $\lambda_0 \mapsto \lambda_1 = \lambda_0 + \Delta \lambda$, the adiabatic energy shell changes, but its enclosed volume remains the same: $\Omega(E_{\Omega_0}(\lambda_1), \lambda_1) = \Omega_0$. 
Each point $q_i$ also changes with $\lambda$, but the volume it bounds remains the same: $S(q_i, \lambda_1) = S(q_i, \lambda_0)$.
The rate at which $q_i$ moves, with respect to variations in $\lambda$, is given by
\begin{equation}
\label{eq:v-total-der-qi}
    v(q_i, \lambda) = \frac{d q_i}{d \lambda} \Big|_S = - \frac{\partial_{\lambda} S}{\partial_q S},
\end{equation}
obtained by setting
\begin{equation}
    d S = \frac{\partial S}{\partial \lambda} d \lambda + \frac{\partial S}{\partial q} d q = 0 .
\end{equation}
As $N \to \infty$,  Eqs.~\eqref{eq:S-qi}, \eqref{eq:v-total-der-qi} define a function 
\begin{align}
\label{eq:app-v-def-with-S}
    v(q, \lambda) = - \frac{\partial_{\lambda} S}{\partial_q S}, \quad   S(q, \lambda) = \int_{q_l(\lambda)}^{q} \overline{p} dq.
\end{align} 
for any $q$ inside the energy shell.
Since this procedure describes the flow of the points $\left\{ q_i \right\}$ in phase space, the function $v(q, \lambda)$ can be considered as a velocity field.

If the potential $U$ has reflection symmetry around the origin for all $\lambda$, i.e.\ $U(q, \lambda) = U(-q, \lambda)$, then so do the energy shells.
It then follows from the above construction that the velocity field is anti-symmetric:
\begin{equation}
\label{eq:app-v-sym}
    v(q, \lambda) = - v(-q, \lambda).
\end{equation}

Now consider the case where the energy shell is not a simple closed loop but is composed of two open curves, as, for example, occurs in the rotation region of the pendulum, see Fig.~\ref{fig:pendulum-phase-space}.
When the pendulum is in rotation motion, it is either rotating clockwise or counter-clockwise, spanning the entire range $q \in [-\pi, \pi)$ cyclically.
In this case, instead of the turning point $q_l(\lambda)$, we set $q_l = -\pi$ for all values of $\lambda$ corresponding to libration, see Eq.~\eqref{eq:pendulum-turning-point-all-E}.

Lastly, if the energy shell forms more than one closed loop, as occurs below the separatrix of the double-well potential (see red curves in Fig.~\ref{fig:shell-evolution-CD}(a)), then Eq.~\eqref{eq:app-v-def-with-S} applies separately to each loop, and $q_l(\lambda)$ denotes the left turning point of the relevant loop.
For the case of the double well, due to the symmetry, it is enough to calculate $v$ for positive $q$ (see Eq.~\eqref{eq:app-v-sym}).

\subsection{Definition of the velocity at the turning points}

Equation \eqref{eq:app-v-def-with-S} does not define the velocity field at the turning points, since $\partial_q S = p$ vanishes there.
The construction in the previous section offers a solution: the first equality in \eqref{eq:v-total-der-qi} can be applied to the turning points $q_{\rm tp}(\lambda)$, the edges of the energy shell $E_{\Omega_0}(\lambda)$, which are determined by
\begin{equation}
    \overline{p}(q_{\rm tp}, \lambda) = \sqrt{2 (E_{\Omega_0}(\lambda) - U(q_{\rm tp}, \lambda))} = 0.
\end{equation}
The above equation yields the relation 
\begin{equation}
    d \overline{p} = \frac{\partial \overline{p}}{\partial q_{\rm tp}} dq_{\rm tp} + \frac{\partial \overline{p}}{\partial \lambda} d \lambda = 0.
\end{equation}
Upon rearranging, we find the velocity field at the turning points (recall the first equality in \eqref{eq:v-total-der-qi})
\begin{equation}
\label{eq:app-v-turning-point-derivative-expression}
    v_{q_{\rm tp}}(\lambda) = \frac{d q_{\rm tp}}{d \lambda} = \frac{\partial_{\lambda} \left( E_{\Omega_0}(\lambda) - U(q_{\rm tp}, \lambda) \right)}{\partial_q U(q_{\rm tp}, \lambda)}.
\end{equation}

\section{Calculation of the velocity field}
\label{sec:calc-veloc-field}

To calculate the velocity field, Eq.~\eqref{eq:v-def}, one needs to calculate the derivative
\begin{align}
\label{eq:v-numerator}
\frac{\partial }{\partial \lambda} \int_{q_{l}(\lambda)}^q \overline{p}(q', \lambda) dq'
&= \int_{q_{l}(\lambda)}^{q} \frac{1}{\overline{p}} \left( \frac{\partial E_{\Omega_0}}{\partial \lambda} - \frac{\partial U}{\partial \lambda}  \right) dq'.
\end{align}
While we do not have an explicit expression for the function $E_{\Omega_0}(\lambda)$, its derivative follows from an identity of partial derivatives,
\begin{equation}
\label{eq:dE-d-lambda-identity}
\frac{\partial E_{\Omega_0}}{\partial \lambda} = 
\left( \frac{\partial E}{\partial \lambda} \right)_{\Omega} = - \frac{\left( \partial_{\lambda} \Omega \right)_E}{\left( \partial_E \Omega \right)_{\lambda}}.
\end{equation}
Thus if the function
\begin{equation}
\label{eq:app-omega-def}
\Omega(E, \lambda) = \oint_{H(q, p, \lambda) = E} p dq
\end{equation}
and its derivatives can be evaluated, then Eq.~\eqref{eq:dE-d-lambda-identity} yields $\partial_{\lambda} E$, which in turn yields the velocity field $v$ (Eq.~\eqref{eq:v-def}).

\section{Double-Well --- Explicit Calculations}
\label{sec:double-well-appendix}

Here we calculate $v(q,\lambda)$ explicitly for the symmetric double-well system given by Eq.~\eqref{eq:double-well}.
In our calculation, it is convenient initially to consider $E$ as an independent parameter.
After solving for $v(q, E, \lambda)$, the argument $E$ is replaced by function $E_{\Omega_0}(\lambda)$, defined by Eq.~\eqref{eq:adiabaticEnergy}, i.e. $v(q, \lambda) = v(q, E_{\Omega_0}(\lambda), \lambda)$.

\subsection{Energy shells and turning points}
\label{sec:dw-energy-shells-and-turning-points}

For all values of $\lambda$, the energy of the separatrix is given by $E_s = 0$.
Negative energies (``below the separatrix") correspond to the particle being trapped in one of the wells, and positive energies (``above the separatrix") correspond to the particle exploring both wells.

The turning points are those where the momentum vanishes. Hence, denoting the roots of $\overline{p}(q)$ by $q_1, q_2$, we can write
\begin{align}
\nonumber
  \overline{p}(q;  q_1, q_2) &= \sqrt{2 \left( E - \frac{q^{4}}{4} + \frac{\lambda^2 q^2}{2} \right)} \\
\label{eq:dw-pbar}
  &= \sqrt{\frac{1}{2} \left( q_2^2 - q^2 \right) \left( q^2 - q_1^2 \right)}.
\end{align}
Solving for $\overline{p}(q, E, \lambda) = 0$ yields
\begin{align}
\label{eq:dw-turning-point-sqrt-1}
q_1(E, \lambda) &= \sqrt{\lambda^2 - \sqrt{\lambda^{4} + 4E}} \\
\label{eq:dw-turning-point-sqrt-2}
q_2(E, \lambda) &= \sqrt{\lambda^2 + \sqrt{\lambda^{4} + 4E}},
\end{align}
where these become a function of the single parameter $\lambda$ by inserting $E = E_{\Omega_0}(\lambda)$.

Below the separatrix, each energy shell is composed of two disjoint ergodic components, which are symmetric with respect to $q=0$, see red loops in Fig.~\ref{fig:shell-evolution-CD}(a).
Each ergodic component has two turning points. The inner and outer turning points are denoted by $\pm q_1$ and $\pm q_2$, respectively (Eqs.~\eqref{eq:dw-turning-point-sqrt-1},~\eqref{eq:dw-turning-point-sqrt-2}).
The turning points mark the boundaries of the energy shell: a point $q$ on the energy shell must satisfy $\vert q\vert \in [q_1, q_2]$.

Above the separatrix, $E>0$, each energy shell is a simple closed loop, see blue loop in Fig.~\ref{fig:shell-evolution-CD}(a).
In terms of the turning points, we find that $q_1^2 < 0$, meaning that $q_1$ is purely imaginary, and $q_2^2>0$.
Thus, for $q$ to be in the energy shell, $q \in [-q_2, q_2]$.

\subsection{The enclosed volume in an energy shell}
\label{sec:enclosed-volume-an}

The volume enclosed by the energy shell $E$, Eq.~\eqref{eq:app-omega-def}, is
\begin{equation}
\label{eq:17}
\Omega(E, \lambda) =  \begin{cases}
                                    4\int_{q_1}^{q_2} \overline{p} dq, & E <0 \\
                                    4\int_{0}^{q_2} \overline{p} dq, & E >0,
                                  \end{cases}
                                \end{equation}
where $\overline{p}$ is given in Eq.~\eqref{eq:pbar-def} and the factor of 4 accounts for the volume enclosed in the regions $q<0$ and $p<0$.

Using elliptic integrals of the first and second kinds,
\begin{align}
\label{eq:elliptic-integrals-1}
E_f(0 \leq \phi \leq \pi / 2, m) &=  \int_{0}^{\phi} \frac{d \theta}{\sqrt{1 - m \sin^2 \theta}} \\
\label{eq:elliptic-integrals-2}
E_e(0 \leq \phi \leq \pi / 2, m) &=  \int_{0}^{\phi} \sqrt{1 - m \sin^2 \theta}d \theta \\
\notag
E_k(m) &=  E_f(\phi = \pi / 2, m) \\
\notag
E_e(m) &=  E_e(\phi = \pi / 2, m),
\end{align}
we obtain the explicit expressions
\begin{widetext}
\begin{align}
\label{eq:dw-omega}
\Omega(E, \lambda) = \frac{4 }{3 \sqrt{2}} \begin{cases}
                      q_2 \left[ \left( q_1^2 + q_2^2 \right) E_e \left( 1 - \frac{q_1^2}{q_2^2} \right) - 2q_1^2 E_k\left( 1 - \frac{q_1^2}{q_2^2} \right) \right] , & E < 0 \\
                     \sqrt{-q_1^2} \left[ \left( q_1^2 + q_2^2 \right) E_e \left( \frac{q_2^2}{q_1^2} \right) + \left( q_2^2 - q_1^2 \right) E_k \left( \frac{q_2^2}{q_1^2} \right) \right], & E > 0.
                    \end{cases}
\end{align}
\end{widetext}

Combining Eq.~\eqref{eq:dE-d-lambda-identity} with the explicit expressions for $q_{1,2}(E, \lambda)$ (Eqs.~\eqref{eq:dw-turning-point-sqrt-1},~\eqref{eq:dw-turning-point-sqrt-2}), yields
\begin{align}
\label{eq:dw-dE-d-lambda-explicit}
\left( \frac{\partial E}{\partial \lambda}  \right)_{\Omega} = \begin{cases}
  - \lambda q_2^2  \frac{E_e \left(1- \frac{q_1^2}{q_2^2} \right)}{E_k \left( 1- \frac{q_1^2}{q_2^2} \right)} , & E < 0 \\
\lambda q_1^2 \left( \frac{E_e \left( \frac{q_2^2}{q_1^2} \right)}{E_k \left( \frac{q_2^2}{q_1^2} \right)} - 1 \right), & E > 0.
\end{cases}
\end{align}

\subsection{The velocity field}
\label{sec:velocity-field-appendix}

Calculating the velocity field, Eq.~\eqref{eq:v-def}, amounts to evaluating the integrals in Eq.~\eqref{eq:v-numerator}.
Defining
\begin{align}
\label{eq:app-A1}
  A_1(q, \lambda) &= \int_{q_{\rm tp}(\lambda)}^q \frac{1}{\overline{p}} dq' \\
\label{eq:app-A2}
  A_2(q, \lambda) &=  \int_{q_{\rm tp}(\lambda)}^q \frac{q'^2}{\overline{p}} dq',
\end{align}
we have
\begin{align}
  \nonumber
v(q, \lambda) &= -\frac{1}{\overline{p}} \frac{\partial }{\partial \lambda} \int_{q_{\rm tp}(\lambda)}^q \overline{p}(q', \lambda) dq' \\
\label{eq:dw-v-in-A12}
&= - \frac{1}{\overline{p}} \left(  \frac{\partial E_{\Omega_0}}{\partial \lambda} A_1 + \lambda A_2 \right).
\end{align}
The lower limit $q_{\rm tp}(E, \lambda)$ of the integrals $A_1, A_2$ in Eqs.~\eqref{eq:app-A1}, \eqref{eq:app-A2}, is taken as 
\begin{equation}
\label{eq:19}
q_{\rm tp}(\lambda) = \begin{cases}
                 q_1, & E_{\Omega_0}(\lambda)<0 \\
                 0, & E_{\Omega_0}(\lambda)>0.
               \end{cases}
\end{equation}
This choice means that we are calculating $v(q\ge 0, \lambda)$ and as the potential $U(q, \lambda)$ is symmetric, we use Eq.~\eqref{eq:app-v-sym}.

The integrals $A_1, A_2$, Eqs.~\eqref{eq:app-A1}, \eqref{eq:app-A2}, can be found in tables  \cite{Friedman_1971_HandbookEllipticIntegrals}
\begin{widetext}
\begin{align}
\label{eq:dw-A1}
A_1 &= \sqrt{2} \begin{cases}
         \frac{1}{q_2}  E_f \left( \sin^{-1} \left( \psi \right) , m \right), & E < 0 \\
         \sqrt{\frac{1}{q_2^2 - q_1^2}} E_f \left(  \sin^{-1} \left( \frac{1}{\psi} \right), \frac{1}{m} \right), & E>0
       \end{cases} \\
\label{eq:dw-A2}
  A_2 &= \sqrt{2} \begin{cases}
            q_2  E_e \left( \sin^{-1} \left( \psi \right) , m \right) -  \frac{\sqrt{(q^2 - q_1^2) (q_2^2 - q^2)}}{q}, & E<0 \\
           -q \sqrt{\frac{q_2^2-q^2}{q^2-q_1^2}} + \sqrt{q_2^2 - q_1^2} E_e \left( \sin^{-1} \left(\frac{1}{\psi}\right) , \frac{1}{m} \right)+  \frac{q_1^2}{\sqrt{q_2^2 - q_1^2}} E_f \left( \sin^{-1} \left(\frac{1}{\psi}\right), \frac{1}{m} \right), & E>0
         \end{cases}\\
    \notag
  \psi & \equiv  \sqrt{\frac{q_2^2 \left( q^2 - q_1^2 \right)}{q^2 \left( q_2^2 - q_1^2 \right)}} = \sqrt{\frac{1 - \frac{q_1^2}{q^2}}{1 - \frac{q_1^2}{q_2^2}}} \\
    \notag
m & \equiv 1 - \frac{q_1^2}{q_2^2}
\end{align}
\end{widetext}
Note that $0 \leq \psi \leq 1$ for $E< 0$ and $0 \leq \psi^{-1} \leq 1$ for $E>0$, so all functions above are well-defined.
Substituting Eqs. \eqref{eq:dw-dE-d-lambda-explicit}, \eqref{eq:dw-A1}, \eqref{eq:dw-A2} into the velocity field, Eq. \eqref{eq:dw-v-in-A12}, we obtain an explicit expression for $v(q, \lambda) = v(q; q_1(\lambda), q_2(\lambda))$.

\subsection{The velocity at the turning points}
\label{sec:app-v-at-turning-points}

The values of the velocity at the turning points are given by Eq.~\eqref{eq:app-v-turning-point-derivative-expression}, but they can be easily calculated using the derivative expression $d q_{\rm tp}/d \lambda$ in Eq.~\eqref{eq:app-v-turning-point-derivative-expression}, for $q_1(E(\lambda),\lambda), q_2(E(\lambda),\lambda)$ in Eqs.~\eqref{eq:dw-turning-point-sqrt-1},~\eqref{eq:dw-turning-point-sqrt-2}:
\begin{align}
\label{eq:v-turning-points-general-formula-explicit}
v_{q_j}(\lambda) &= \frac{d q_j}{d \lambda} = \frac{\partial q_j}{\partial \lambda} + \frac{\partial q_j}{\partial E} \frac{\partial E}{\partial \lambda}, \quad j=1,2.
\end{align}
The calculation yields
\begin{align}
\label{eq:v-q_1-below}
v_{q_1} &= \frac{2 \lambda}{q_1 \left(1 - \frac{q_1^2}{q_2^2} \right)} \left(  \frac{E_e \left( 1 - \frac{q_1^2}{q_2^2} \right)}{E_k \left( 1 - \frac{q_1^2}{q_2^2} \right)} - \frac{q_1^2}{q_2^2}\right), \quad E<0, \\
\notag
v_{q_2} &= \begin{cases}
             \frac{2 \lambda q_2}{q_2^2 - q_1^2} \left( 1-  \frac{E_e \left( 1 - \frac{q_1^2}{q_2^2} \right)}{E_k \left( 1 - \frac{q_1^2}{q_2^2} \right)} \right), \quad E < 0 \\
             \frac{2 \lambda}{q_2 \left( q_2^2 - q_1^2 \right)} \left[ q_1^2 \frac{E_e \left( \frac{q_2^2}{q_1^2} \right)}{ E_k \left( \frac{q_2^2}{q_1^2} \right)} + q_2^2 - q_1^2 \right], \quad E>0.
           \end{cases}
\end{align}
On the separatrix, $q_1 = 0$, hence $v_{q_1} \to \infty$ there.
It is clear that the source of the divergence comes only from the factor $1 / q_1$.
Indeed we find that $\lim_{q_1 \to 0} v_{q_1} = \infty$, but $\lim_{q_1 \to 0} q_1 v_{q_1} < \infty$.

\subsection{The acceleration field}
\label{sec:app-dw-acc-field}

The forces required for FF driving are given by the acceleration field (see Eq. (8) in Ref.~\cite{Subasi_2017_FastForwardClassical})
\begin{align}
\nonumber
\tilde{a}(q, t) &= \frac{\partial \tilde{v}}{\partial q} \tilde{v} + \frac{\partial \tilde{v}}{\partial t} \\
\nonumber
&= v \dot{\lambda} \frac{\partial v}{\partial q} \dot{\lambda} + \frac{\partial v \dot{\lambda} }{\partial t} \\
\nonumber
&= v  \frac{\partial v}{\partial q} \left( \dot{\lambda} \right)^2 + \frac{\partial v  }{\partial t} \dot{\lambda} + v \ddot{\lambda} \\
\nonumber
&= \left( v  \frac{\partial v}{\partial q}  + \frac{\partial v  }{\partial \lambda} \right) \dot{\lambda}^2  + v \ddot{\lambda} \\
& \equiv a(q, \lambda) \dot{\lambda}^2 + v(q, \lambda) \ddot{\lambda},
\label{eq:a-field}
\end{align}
where we used the decomposition $\tilde{v}(q, t) = v(q, \lambda(t)) \dot{\lambda}(t)$, see Eq.~\eqref{eq:H_CD}.
As $v(q, \lambda)$ is given in Eqs.~\eqref{eq:dw-v-in-A12}, \eqref{eq:dw-A1}, \eqref{eq:dw-A2}, the derivatives $\partial_q v, \partial_{\lambda} v$ can be readily calculated to yield $a(q, \lambda)$ and in turn $\tilde{a}(q, t)$.
Thus, the FF forces are explicitly calculated in terms of the function $E_{\Omega_0}(\lambda)$.

As $v(q, \lambda)$ and its derivatives appear in $a(q, \lambda)$, the acceleration field also diverges when crossing a separatrix. The idea of regularizing the divergence by incorporating an inflection point in the protocol $\lambda(t)$ (Eq. \ref{eq:lambda-t}) when crossing the separatix works for the acceleration field as well. Namely, even though $a(q, \lambda), v(q, \lambda)$ diverge at the separatrix, a proper choice of $\lambda(t)$ regularizes this divergence, making $\tilde{a}(q,t)$ finite there.

\section{Pendulum --- Explicit Calculations}
\label{sec:pendulum-appendix}

Here we provide explicit calculations of the velocity field for the pendulum 
\begin{equation}
\label{eq:H-pendulum}
    H(q, p, \lambda) = \frac{p^2}{2} + \lambda \left( 1 - \cos q \right),
\end{equation}
where $q \in \left( -\pi, \pi \right)$ and \(\lambda\) is gravity.
The procedure is the same as for the double-well system in Sec.~\ref{sec:double-well-appendix}.

\begin{figure}[htbp]
  \centering
  \includegraphics[width=0.99\columnwidth]{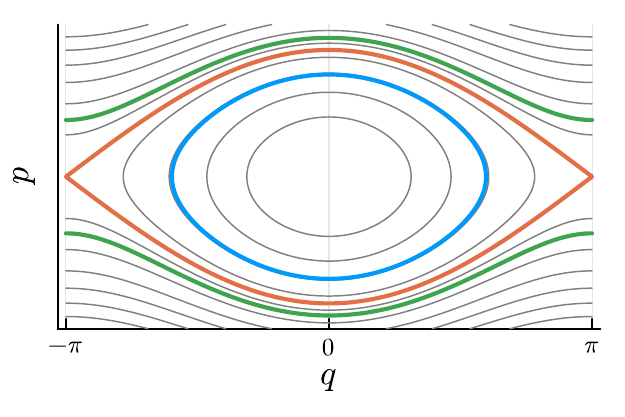}
  \caption{Phase space of the pendulum, Eq.~\eqref{eq:H-pendulum}. Three energy shells are highlighted: below the separatrix (blue), on the separatrix (orange), and above the separatrix (green). Note that above the separatrix, the energy shell is composed of two disjoint ergodic components that have no turning points corresponding to clockwise and counter-clockwise rotations.}
  \label{fig:pendulum-phase-space}
\end{figure}

The pendulum has a separatrix at $E_s=2 \lambda$. 
The libration region corresponds to energies in the range $E \in[0, 2 \lambda)$, whereas the rotation region corresponds to energies in the range $E \in (2 \lambda, \infty)$, see Fig.~\ref{fig:pendulum-phase-space}.

The rotation region has no turning points. 
In fact, an energy shell in the rotation region is composed of two disjoint ergodic components corresponding to clockwise and counter-clockwise directions. 
As the range of the coordinate $q$ is $[- \pi, \pi)$, we denote the turning points $\pm q_1(E, \lambda)$ by
\begin{equation}
\label{eq:pendulum-turning-point-all-E}
q_1(E, \lambda) = \begin{cases}
  2 \sin^{-1} \left( \sqrt{\frac{E}{2\lambda}} \right), & 0 \leq E < 2 \lambda \\
 \pi, &  2 \lambda < E.
\end{cases}
\end{equation}

The enclosed volume in an energy shell $E$, for a specific value of the control parameter $\lambda$, is given by
\begin{align}
\label{eq:pendulum-enclosed-volume}
&\Omega(E, \lambda) = \oint p dq =  4\int_{0}^{q_1} \overline{p} dq \\
&\overline{p} = \sqrt{2 \left( E - \lambda \left( 1 - \cos q \right) \right)} = \sqrt{2E \left( 1 - \frac{2\lambda}{E} \sin^2 \left(\frac{q}{2} \right) \right)}.
\end{align}
Explicit calculation yields
\begin{equation}
\label{eq:pend-omega}
    \Omega(E, \lambda) = 8 \sqrt{2 E} E_e \left( \frac{q_1(E, \lambda)}{2} ; \frac{2 \lambda}{E} \right),
\end{equation}
where $E_e(\phi, m)$ is the elliptic integral of the second kind, see Eq.~\eqref{eq:elliptic-integrals-2}.

Using Eq.~\eqref{eq:dE-d-lambda-identity}, with the explicit expressions for $q_{1}(E, \lambda)$, Eq.~\eqref{eq:pendulum-turning-point-all-E}, and $\Omega(E, \lambda)$, Eq.~\eqref{eq:pend-omega}, while treating $E, \lambda$ as independent variables, yields
\begin{align}
\label{eq:pendulum-dE-d-lambda}
\left( \frac{\partial E}{\partial \lambda}  \right)_{\Omega} = 
\begin{cases}
     2 - 2 \frac{E_e \left( \frac{q_1}{2} ; \frac{2 \lambda}{E} \right)}{E_f \left( \frac{q_1}{2} ; \frac{2 \lambda}{E} \right)} , & E < 2 \lambda \\
    \frac{E}{\lambda} \left( 1 - \frac{E_e \left( \frac{q_1}{2} ; \frac{2 \lambda}{E} \right)}{E_f \left( \frac{q_1}{2} ; \frac{2 \lambda}{E} \right)} \right), & E > 2 \lambda.
\end{cases}
\end{align}

Now, all ingredients are in place to calculate $v(q,\lambda)$ explicitly.
Using Eq.~\eqref{eq:v-numerator} and Eq.~\eqref{eq:pendulum-dE-d-lambda}, we find
\begin{align}
   v &= - \frac{1}{\overline{p}} \int_{-q_1}^{q} \frac{1}{\overline{p}} \left( \frac{\partial E_{\Omega_0}}{\partial \lambda} - \frac{\partial U}{\partial \lambda}  \right) dq' \\
   &=  - \frac{1}{\overline{p}} \frac{\sqrt{2 E}}{\lambda} \left[ E_e \left( \frac{q}{2}; \frac{2 \lambda}{E} \right) - \frac{E_e \left( \frac{q_1}{2} ; \frac{2 \lambda}{E} \right)}{E_f \left(\frac{q_1}{2} ; \frac{2 \lambda}{E} \right)} E_f \left( \frac{q}{2}; \frac{2 \lambda}{E} \right) \right],
\end{align}
where $-q_1 \leq q \leq q_1$, and $q_1$ is given in Eq.~\eqref{eq:pendulum-turning-point-all-E}.

Next, we find the velocity field at the turning points $v_{q_1}$.
Recall that $q_1(E, \lambda)$, Eq.~\eqref{eq:pendulum-turning-point-all-E},  depends on $E, \lambda$ only below the separatrix, namely for $E < 2 \lambda$. 
Using Eq.~\eqref{eq:v-turning-points-general-formula-explicit} for $q_1(E, \lambda)$ in Eq.~\eqref{eq:pendulum-turning-point-all-E} and using Eq.~\eqref{eq:pendulum-dE-d-lambda}, we find, denoting $x = E / (2\lambda) < 1$
\begin{equation}
\label{eq:pend-v-turning-point}
    v_{q_1} = \frac{1}{\lambda \sqrt{x(1-x)}} \left( (1-x) - \frac{E_e(x)}{E_f(x)} \right).
\end{equation}
Approaching the separatrix corresponds to $x \to 1$.
The explicit form of $v_{q_1}$ shows that $\lim_{x\to 1} v_{q_1} = - \infty$ due to the second term in Eq.~\eqref{eq:pend-v-turning-point}.
Since the divergence is solely due to the term $\sqrt{1-x}$ in the denominator, it is clear that $|\lim_{x \to 1} \sqrt{1-x} v_{q_1}| < \infty$.

To get the forces of FF driving, the same calculation as in Sec.~\ref{sec:app-dw-acc-field} is followed.

\end{document}